\documentclass[jcp, floatfix, nobibnotes, reprint, superscriptaddress]{revtex4-1}
\usepackage{docs}%
\usepackage{siunitx}
\usepackage{amsmath}
\usepackage{booktabs}
\usepackage{amssymb}
\DeclareSIUnit[number-unit-product = {\,}]
\cal{cal}
\DeclareSIUnit\kcal{\kilo\cal}
\DeclareSIUnit[number-unit-product = {\,}]
\Btu{Btu}
\DeclareSIUnit[number-unit-product = {\,}]
\Fahr{\degree F}
\DeclareSIUnit[number-unit-product = {\,}]
\usepackage{bm}%

\usepackage{graphicx}
\usepackage[utf8]{inputenc}
\usepackage{tabularx} 

\usepackage{dcolumn}
\usepackage{epstopdf}
\usepackage{afterpage}
\usepackage{setspace}
\usepackage{multirow}
\usepackage{dsfont}
\usepackage{sidecap}
\usepackage{epstopdf}
\usepackage{braket}
\usepackage{siunitx} 
\usepackage{color, colortbl}

\definecolor{Gray}{gray}{0.9}

\AppendGraphicsExtensions{.gif}

\DeclareMathAlphabet\mathbfcal{OMS}{cmsy}{b}{n}

\setcitestyle{super}

\begin{document}

\title{Quantum mechanical dataset of 836k neutral closed shell molecules with upto 5 heavy atoms from CNOFSiPSClBr}
\author{Danish Khan}
\affiliation{Chemical Physics Theory Group, Department of Chemistry, University of Toronto,
St. George Campus, Toronto, ON, Canada}
\affiliation{Vector Institute for Artificial Intelligence, Toronto, ON, M5S 1M1, Canada}

\author{Anouar Benali}
\affiliation{Computational Science Division, Argonne National Laboratory, Argonne, Illinois 60439, United States}

\author{Scott Y. H. Kim}
\affiliation{Chemical Physics Theory Group, Department of Chemistry, University of Toronto,
St. George Campus, Toronto, ON, Canada}
\affiliation{Vector Institute for Artificial Intelligence, Toronto, ON, M5S 1M1, Canada}

\author{Guido Falk von Rudorff}
\affiliation{Institute of Chemistry, University of Kassel, 34109 Kassel, Germany}
\affiliation{Center for Interdisciplinary Nanostructure Science and Technology (CINSaT), 34132 Kassel, Germany}

\author{O. Anatole von Lilienfeld}
\email{anatole.vonlilienfeld@utoronto.ca}
\affiliation{Chemical Physics Theory Group, Department of Chemistry, University of Toronto, St. George Campus, Toronto, ON, Canada}
\affiliation{Department of Materials Science and Engineering, University of Toronto, St. George Campus, Toronto, ON, Canada}
\affiliation{Vector Institute for Artificial Intelligence, Toronto, ON, M5S 1M1, Canada}
\affiliation{ML Group, Technische Universit\"at Berlin and Institute for the Foundations of Learning and Data, 10587 Berlin, Germany}
\affiliation{Berlin Institute for the Foundations of Learning and Data, 10587 Berlin, Germany}
\affiliation{Department of Physics, University of Toronto, St. George Campus, Toronto, ON, Canada}
\affiliation{Acceleration Consortium, University of Toronto, Toronto, ON}
\begin{abstract}
We introduce the Vector-QM24 (VQM24) dataset comprehensively covering all possible neutral closed-shell small organic and inorganic molecules with up to five heavy (\textit{p}-block) atoms: C, N, O, F, Si, P, S, Cl, Br. 
All valid stoichiometries, Lewis-rule-consistent graphs, and stable conformers (identified via GFN2-xTB) were enumerated combinatorially, yielding 577k conformational isomers spanning 258k constitutional isomers and 5,599 unique stoichiometries. DFT ($\omega$B97X-D3/cc-pVDZ) optimizations were performed for all, and diffusion quantum Monte Carlo (DMC@PBE0(ccECP/cc-pVQZ)) energies are provided for 10,793 lowest-energy conformers with up to 4 heavy atoms. 
VQM24 includes structures, vibrational modes, rotational constants, thermodynamic properties (Gibbs free energies, enthalpies, ZPVEs, entropies, heat capacities), and electronic properties such as atomization, electron interaction, exchange-correlation, dispersion energies, multipole moments (dipole to hexadecapole), alchemical potentials, Mulliken charges, and wavefunctions. 
Machine learning models of atomization energies on this dataset reveal significantly higher complexity than QM9, with none achieving chemical accuracy. VQM24 offers a rigorous, high-fidelity benchmark for evaluating quantum machine learning models.
\end{abstract}
\maketitle
\subsection{Introduction}
High quality quantum mechanical datasets of molecular properties are a  primary requirement for developing approximate physics and statistics based models to enhance the navigation of chemical compound space (CCS).
Numerous datasets focusing on distinct, chemically relevant subspaces have paved the way for systematic and quantitative exploration of CCS such as in Refs.~\cite{qm7,qm7b,qm9,gdml,anders_gradients_role,QMugs,geom,multixcqm9,ani1_dataset,ANI-1,ani1-x,qm7-x,egp,pubchemqc,SPICE}. 
Most of the quantum mechanical (QM) datasets such as QM7\cite{qm7}, QM9\cite{qm9}, ANI\cite{ANI-1}, QMrxn~\cite{qmrxn}, QMugs\cite{QMugs}, PubChemQC\cite{pubchemqc} are derived from string based lists of compounds from the GDB\cite{GDB13,GDB17}, ChEMBL\cite{chembl}, PubChem\cite{pubchem} databases while datasets like revQM9~\cite{apbe0}, GEOM\cite{geom}, MultiXC-QM9\cite{multixcqm9}, G4MP2-QM9~\cite{g4mp2qm9}, QMspin~\cite{qmspin}, QM-sym~\cite{qm-sym,qm-symex}, ANI-1x\cite{ani1-x}, QM7-X\cite{qm7-x} correspond to extensions.
The effectiveness of ML models relies on complete representativeness and accuracy of the relevant reference data. 
Unfortunately, and due to the combinatorial scaling of number of possible stable compounds with size and composition~\cite{von2013first}, they are typically incomplete and consequently introduce considerable bias in machine learning (ML) models trained and assessed on them. 
Furthermore, while Density Functional Theory (DFT) has been key to the development of highly accurate and efficient ML models over the past decade\cite{central_role_dft}, there is still a lack of data exhaustively covering specific regions of chemical space at higher QM levels.
Note that even for the simplest stoichiometries and smallest subsets of graphs, exhaustive lists of quantum properties are lacking. 
The primary reason behind this is the combinatorially intractable nature of the problem as the number of atoms and unique chemical elements grows.
Nevertheless, it is important to systematically fill these gaps since limited chemical diversity limits the generalizability of ML models, as was pointed out recently\cite{Glavatskikh2019}.
Furthermore, exhaustive coverage of chemical spaces spanned by the smallest systems is key for ML models since locality can be exploited to achieve scalable statistical models of physical  properties\cite{amons_slatm,gems,bing2023,apbe0,abasis}.\\

Here, we tackle this task by reporting VECTOR-QM24 (VQM24), a diverse and comprehensive dataset of - small organic and inorganic molecules calculated at the $\omega$B97X-D3/cc-pVDZ level of theory\cite{wb97xd,dunning1989gaussian,wB97X,d3}.
VQM24 comprises 5,599 unique stoichiometries, corresponding to 258,242 distinct molecular graphs and constitutional isomers. 
Ground state structures of these constitutional isomers were used to obtain 577,705 additional conformers, leading to a grand total of 835,947 molecular structures within the dataset.
More specifically, this dataset has been generated by first
evaluating all possible Lewis structures (according to {\tt SURGE}~\cite{surge_graph_gen_2022}) for molecules consisting of up to five heavy atoms drawn from C, N, O, F, Si, P, S, Cl, Br with their most frequent valencies followed by saturation with hydrogens to obtain neutral closed-shell combinations. 
Thereafter, conformational isomers were generated for all graphs, 
using GFN2-xTB~\cite{gfn2xtb}, and subsequently relaxed using
density functional theory ($\omega$B97X-D3/cc-pVDZ). 
For all 835,947 converged molecules (post DFT optimization), we provide the corresponding optimized structures, an extensive list of thermal properties (internal, atomization, Gibbs free and zero point vibrational energies; enthalpy, entropy, heat capacities and rotational constants) along with vibrational modes and frequencies, electronic properties (electron repulsion, exchange-correlation, dispersion and moleculuar orbital energies; HOMO-LUMO gaps, electrostatic potentials at nuclei, dipole, quadrupole, octupole, hexadecapole moments)  and wavefunctions.
Goind beyond DFT, we also report diffusion quantum Monte Carlo (DMC) energies, converged to sub milli-Hartree statistical uncertainty, for the smaller sub-set of 10,793 energetically lowest lying conformers of molecules composed of only up to 4 heavy atoms. 
To the best of our knowledge, this constitutes the largest quantum Monte Carlo (QMC) dataset in chemical space reported yet. 
The molecules included in this dataset also constitute an overlapping as well as complementary set of the atoms-in-molecules (amons)\cite{amons_slatm} dictionary that represents the local chemistries encoded by the GDB and ZINC~\cite{huang2020dictionary} lists. 
\begin{figure*}[!h]
    \centering
    \includegraphics[width=0.6\linewidth]{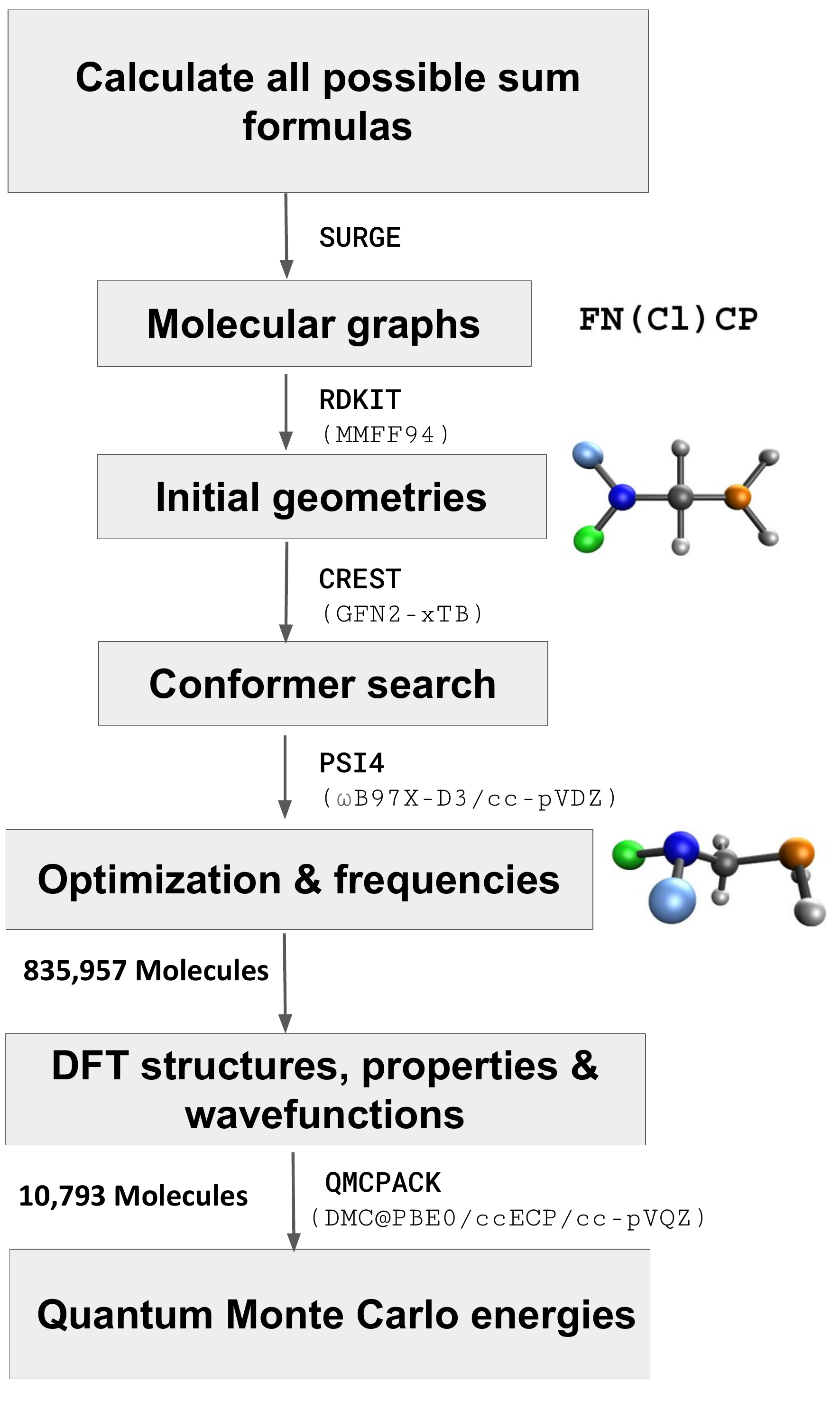}
    \caption{Workflow used to generate the VQM24 dataset. All possible stoichiometries were first calculated by choosing all combinations of up to 5 heavy atoms (non-Hydrogen) and saturating them with hydrogens to satisfy the valencies via integer partitioning. Heavy atoms included along with their valencies are reported in Table~\ref{tab:vqm24_stoich}. For each stoichiometry, all possible graphs as identified by the \texttt{SURGE}\cite{surge_graph_gen_2022} package were evaluated. \texttt{RDkit}\cite{landrum2013rdkit} was then used to generate initial geometries which were first optimized at the \texttt{GFN2-xTB}\cite{gfn2xtb} level of theory, followed by a conformer search using \texttt{CREST}\cite{Crest_2020}. All conformers identified at the xTB level of theory were then optimized with DFT ($\omega$B97X-D3/cc-pVDZ\cite{wB97X,dunning1989gaussian,d3}) using \texttt{PSI4}\cite{psi4_2020}, followed by frequency calculations to identify the saddle point orders. For a smaller subset of the most stable conformers with up to 4 heavy atoms, subsequent diffusion quantum Monte Carlo (DMC) calculations were performed using \texttt{QMCPACK}\cite{kim2018qmcpack,kent2020qmcpack} along with nodal surfaces obtained at the PBE0/ccECP/cc-pVQZ\cite{ccecp_1,ccecp_2,PBE01,PBE02} level of theory using \texttt{PySCF}~\cite{Pyscf2020}.}
    \label{fig:vqm_workflow}
\end{figure*}
\\
Figure~\ref{fig:vqm_workflow} summarizes the entire procedure used to generate the dataset which is explained in the next section.
\section{Methods}
\subsection{Structure generation}
\label{sec: structure_generation}
\begin{table*}[!h]
    \centering
    \begin{tabular}{|l|l|}
        \hline
        Element & Valencies \\
        \hline
        C & 4 \\
        \hline
        N & 3, 5 (${\rm N_{X}}$) \\
        \hline
        O & 2 \\
        \hline
        F & 1 \\
        \hline
        Si & 4 \\
        \hline
        P & 3, 5 (${\rm P_{X}}$) \\
        \hline
        S & 2, 4 (${\rm S_{X}}$), 6 (${\rm S_{Y}}$) \\
        \hline
        Cl & 1 \\
        \hline
        Br & 1 \\
        \hline
    \end{tabular}
    \caption{Chemical elements and their corresponding valencies included in the VQM24 dataset.}
    \label{tab:vqm24_stoich}
\end{table*}

In order to generate the structures for VQM24, all combinatorially possible sum formulas were calculated from molecules with up to five heavy atoms, drawn from the list of the following chemical elements: 
C, N, ${\rm N_{X}}$, O, F, Si, P, ${\rm P_{X}}$, S, ${\rm S_{X}}$, ${\rm S_{Y}}$, Cl, Br (lower index syntax is in line with {\tt SURGE}~\cite{surge_graph_gen_2022} and correspond to less common valencies, as also listed in Table~\ref{tab:vqm24_stoich}).
This was done by generating all possible combinations for up to 5 selected elements (not counting hydrogen) using the native Python package {\tt itertools}. 
Since there are 13 possible heavy elements, the number of combinations containing $n$-heavy atoms is equal to $\frac{(12+n)!}{12!n!} = $ 13, 91, 455, 1820, and 6188 for $n = $1, 2, 3, 4, and 5, respectively.
The possible number of hydrogen atoms, $n_{\rm H}$, for any chosen combination of heavy atoms is then given by the following integer partitioning problem:
\begin{align}
    &n_{\rm H} = v - 2n_{1} - 4n_{2} - 6n_{3}
    \\
    &0 \leq n_{1} \leq \frac{v}{2}, ~ 0 \leq n_{2} \leq \frac{v}{4}, ~ 0 \leq n_{3} \leq \frac{v}{6} \nonumber
\end{align}
where $n_{1}$, $n_{2}$, $n_{3}$ are integers 
respectively denoting the number of single , double and triple bonds in the system, while $v$ denotes the total valency of the system (sum over atomic valencies of each element as displayed in Table~\ref{tab:vqm24_stoich}).
Note that this procedure also accounts for stoichiometries consistent with ring closures. 
As concrete illustrations, consider methylamine (CH$_3$NH$_2$): carbon (valency 4) plus nitrogen (valency 3) gives $v=7$. 
With one C–N single bond ($n_{1}=1$, $n_{2}=n_{3}=0$), Eq. (1) yields $n_{\rm H} = 7 - 2\!\times\!1 = 5$, 
producing CH$_3$NH$_2$.  Likewise, for ethene (C$_2$H$_4$), two carbons ($v=4+4=8$) with one C=C double bond ($n_{2}=1$, $n_{1}=n_{3}=0$) give $n_{\rm H} = 8 - 4\!\times\!1 = 4$,
yielding C$_2$H$_4\,$.
For Benzene (C$_6$H$_6$): six C atoms ($v=6\times4=24$) with three C–C single bonds and three C=C double bonds ($n_{1}=3$, $n_{2}=3$, $n_{3}=0$) $n_{\rm H} = 24 - 2\times3 - 4\times3 = 6$,
yielding C$_6$H$_6$ and capturing the ring closure topology automatically.

For the calculated sum formulas, molecular graphs were generated using \texttt{SURGE}.\cite{surge_graph_gen_2022} 
The graphs were then converted to geometries with MMFF94 as implemented in \texttt{RDKit}\cite{landrum2013rdkit}, 
which were then optimized initially using the \texttt{GFN2-xTB}\cite{gfn2xtb} semi-empirical method. 
Following this, a conformer search was conducted with \texttt{Crest}\cite{Crest_2020}, 
and all conformers were added to the dataset. 
This workflow resulted in $\sim$1.1M geometries.

\subsection{DFT optimization and calculations}
Subsequently, we optimized all geometries with DFT using the $\omega$B97X-D3/cc-pVDZ
level of theory\cite{wB97X,dunning1989gaussian,d3}.
This functional and dispersion correction combination was selected due to its excellent performance in main-group thermochemistry, kinetics, and noncovalent interaction benchmarks such as GMTKN55\cite{gmtkn55}, where it ranks among the most accurate methods—excluding more computationally demanding double hybrid functionals. 
Importantly, our choice aligns with other widely used datasets such as ANI-1\cite{ANI-1}, ANI-1x\cite{ani1-x}, OrbNet Denali\cite{christensen2021orbnet}, QMugs\cite{QMugs}, SPICE\cite{SPICE}, and MultiXC-QM9\cite{multixcqm9}, all of which report values obtained from the $\omega$B97X functional (or its variants) with a double-zeta basis set. 
This consistency ensures that ML models trained on VQM24 can be readily integrated with models developed on related datasets, facilitating broader interoperability and transfer learning across datasets.
\\
The \texttt{Gaussian Tight} convergence criteria (as implemented in \texttt{PSI4}\cite{psi4_2020}) and
 density fitting (for computational efficiency) were employed  in all calculations using the cc-pVDZ-JKFIT\cite{jkfit} auxiliary basis set.
 All DFT calculations were conducted using the \texttt{PSI4} software package (version 1.7).\cite{psi4_2020}
 The optimization was performed in three passes.
 In the first pass the default settings in \texttt{PSI4} for geometry optimizations were used (DIIS method\cite{diis} for SCF and RFO for geometry optimization in redundant internal coordinates\cite{rfo_psi4}) with a maximum of 100 optimization steps.
 The molecules that did not converge entered the second pass in which 2nd order SCF (using the keyword \texttt{SOSCF}) convergence method employing the full Newton step was used along with ultrafine Lebedev-Treutler\cite{treutler1995,Lebedev1999AQF} exchange-correlation integration grid (590 spherical, 99 radial points) and a maximum of 100 geometry optimization steps.
 In the third pass, full Hessian evaluations were performed for the initial geometry and every 20\textit{th} geometry optimization step afterwards with a maximum of 50 steps.
 The optimization was performed in Cartesian coordinates only, in  conjunction with the settings employed in the 2nd pass.
 Those molecules that did not converge after all three passes were left unconverged, and have not been reported.
 Following this procedure, we obtained a grand-total of 835,947 converged molecules, and 262,542 that did not converge.
 
\begin{figure*}[!h]
    \centering
    \includegraphics[width=\linewidth]{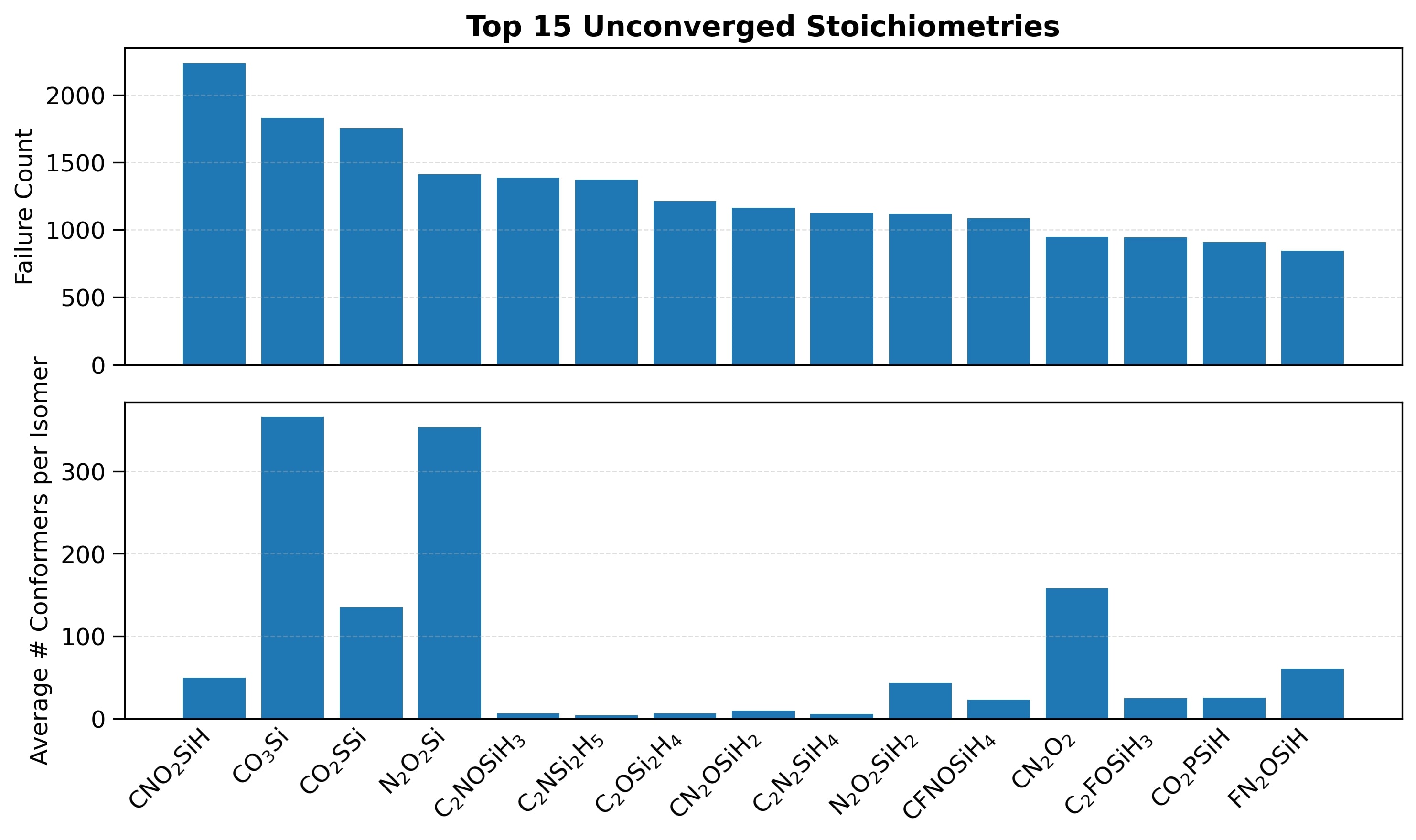}
    \caption{Analysis of DFT geometry-optimization failures.  
(A) (Top) Bar chart of the fifteen stoichiometries with the highest number of unconverged $\omega$B97X‑D3/cc‑pVDZ optimizations. Stoichiometries are ordered by descending failure count.  
(B) (Bottom) Conformer diversity for these stoichiometries, quantified as the mean number of unique CREST~\cite{Crest_2020}-generated (GFN2‑xTB~\cite{gfn2xtb}) conformers per constitutional isomer.}
    \label{fig:vqm_fig3}
\end{figure*}
 Figure~\ref{fig:vqm_fig3} shows an analysis of the most common stoichiometries that failed the DFT geometry optimizations.
 Panel (A) ranks the fifteen stoichiometries with the highest failure counts—fourteen of which contain silicon—highlighting a clear silicon bias in the convergence failures. 
 This is consistent with previous reports that xTB‑based conformer generation can struggle with silicon‑containing compounds, often producing unreliable starting geometries or poor convergence behavior~\cite{silicon_xtb_benchmark,silicon_xtb_improving}. 
 This is also evident from panel (B) which shows the mean number of CREST~\cite{Crest_2020}‑generated (GFN2‑xTB~\cite{gfn2xtb}) conformers per constitutional isomer for these stoichiometries, which far exceeds the dataset average of ~3 conformers. 
 \\
 The relaxation of converged systems was subsequently followed by vibrational frequency calculations at the same level of theory to identify the saddle point orders of the geometries.
 In total we found 784,875 molecules to have converged to a local minimum and 51,072 to saddle points.
 All molecules have been included in the dataset with the minimum geometries and saddle points stored as separate datasets (see Data Records below).
\begin{table*}[!h]
    \centering
    \begin{tabular}{|c|c|c|c|}
        \hline
        \textbf{Heavy atoms ($N$)} & \textbf{Stoichiometries} & \textbf{Graphs} & \textbf{Geometries} \\
        \hline
        1 & 9    & 9      & 9      \\
        \hline
        2 & 69   & 69     & 81     \\
        \hline
        3 & 367  & 766    & 1,287  \\
        \hline
        4 & 1,321 & 10,992 & 29,581 \\
        \hline
        5 & 3,793 & 246,406 & 753,917 \\
        \hline
    \end{tabular}
    \caption{Counts of unique stoichiometries, graphs (constitutional isomers), and geometries (constitutional isomers and conformers) in VQM24, binned by heavy‐atom count.
    }
    \label{tab:unique_counts}
\end{table*}
\begin{figure*}[!h]
    \centering
    \includegraphics[width=0.8\linewidth]{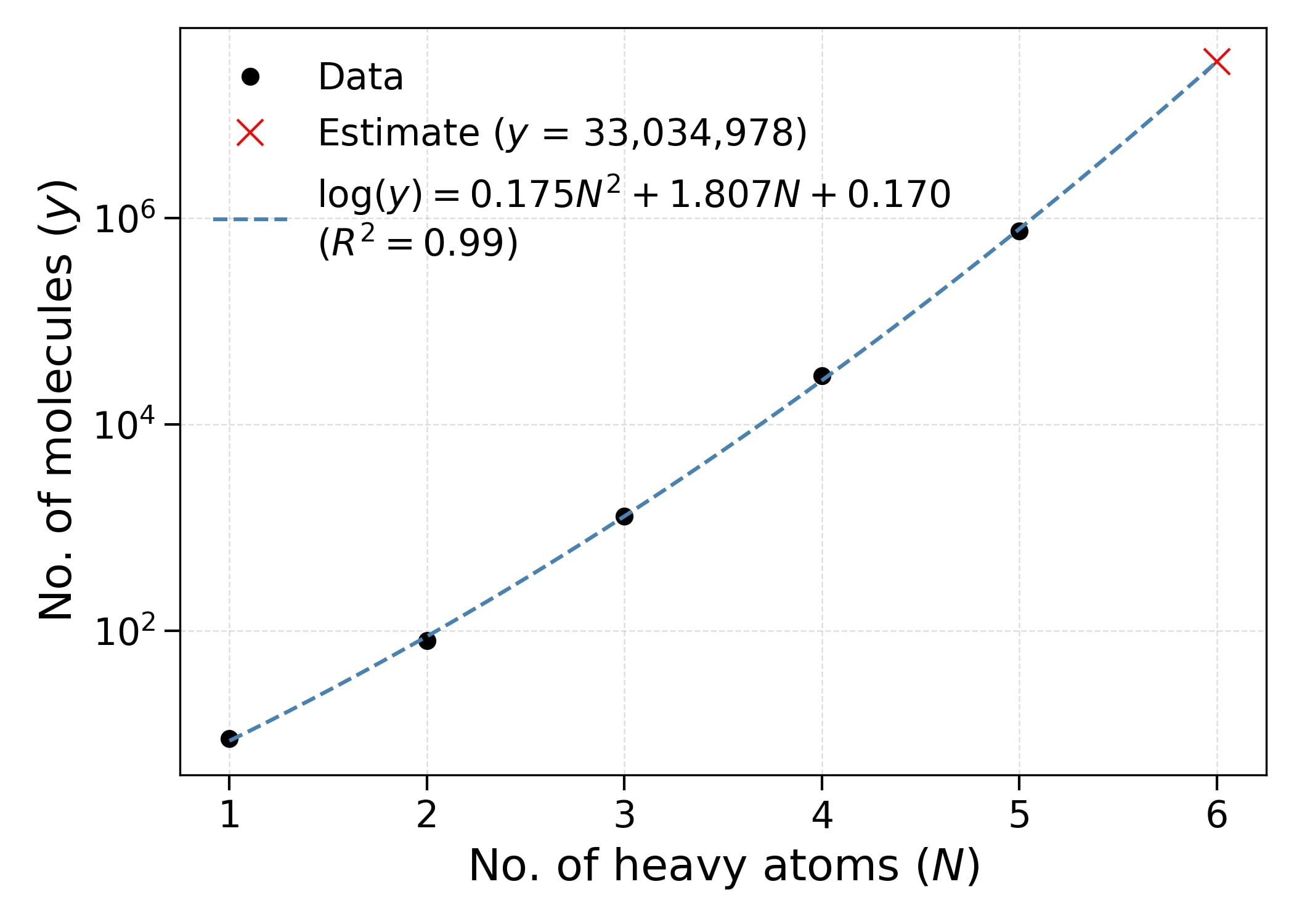}
    \caption{Growth in the number of molecules ($y$) in the VQM24 dataset as a function of heavy atom count ($N$). Each black marker corresponds to the total number of molecular geometries (including conformers) reported in the last column of Table~\ref{tab:unique_counts}. 
A quadratic fit of the form $\log(y) = aN^2 + bN + c$ captures the observed combinatorial scaling, with an excellent correlation coefficient of $R^2 = 0.99$. The red cross indicates the extrapolated estimate for $N = 6$, predicting approximately 33 million distinct molecular geometries.}
    \label{fig:vqm_fig2}
\end{figure*}
 Table~\ref{tab:unique_counts} summarizes the number of unique stoichiometries, constitutional isomers, and geometries geometries (including conformers), grouped by the number of heavy atoms (excluding hydrogens) per molecule, based on the 784,875 minimum-energy geometries reported. 
 Figure~\ref{fig:vqm_fig2} illustrates the growth of the total number of molecules (corresponding to the last column of Table~\ref{tab:unique_counts}) obtained through our procedure.
 The dataset exhibits clear combinatorial scaling, with the projected number of molecules exceeding 30 million upon inclusion of species containing six heavy atoms.
\\
\subsection{Diffusion Monte Carlo}
Quantum Monte Carlo (QMC) techniques are methods that stochastically solve the many-body Schrödinger equation. By explicitly including many-body electronic interactions, these methods achieve mathematical rigor and, in principle, can resolve the Schrödinger equation exactly. However, practical applications require some approximations to maintain computational feasibility, although most of these are controlled and can be rigorously extrapolated at a computational cost. With the proliferation of high-performance computers reaching hundreds of petaflops and the recent deployment of exascale machines, (Summit at Oak Ridge National Laboratory and Aurora at Argonne National Laboratory), QMC methods are poised to significantly take advantage of this computational power; by utilizing stochastic numerical sampling, where samples are evaluated independently, QMC methods achieve embarrassingly parallel processing, enhancing their efficiency for high-performance computing.\\
While  many variations exist, recent years have  seen significant theoretical, algorithmic, and computational advances, particularly in Diffusion Monte Carlo (DMC).
Using a projector or Green's function based approach, DMC solves the Schrödinger equation in an imaginary time $\tau=it$. This ensures that any initial state $|\psi\rangle$, not orthogonal to the ground state $|\phi_0\rangle$, will converge to the ground state in a long time limit. During this process, components corresponding to excited states diminish exponentially, ultimately yielding the true ground state.  
 \begin{equation}
\lim_{\tau \rightarrow \infty} \Psi(\textbf{R},\tau)=c_0 e^{-\epsilon_0\tau}\phi_0(\textbf{R})
\label{EQ:decay}
\end{equation}
The introduction of a constant energy offset, $E_{T}=\epsilon_0$, stabilizes the long-time behavior of the system and keeps it finite. The imaginary time Schr\"odinger equation then resembles a diffusion equation given by:
\begin{equation}\label{eq:Boson}
-\frac{\delta \Psi(\textbf{R},\tau)}{\delta \tau}=\left[\sum_{i=1}^{N}-\frac{1}{2}\nabla^{2}_{i}\Psi(\textbf{R},\tau) \right] + \left[V(\textbf{R})-E_T\right]\Psi(\textbf{R},\tau)
\end{equation}
The first term captures the diffusion of particles, while the second term is a branching term dependent on the potential capturing the change in the density of these particles. The potential $V(\textbf{R})$ in Coulombic systems is unbounded, which may cause the rate term $\left(V(\textbf{R}) - E_T\right)$ to diverge. This could lead to considerable fluctuations in particle density and cause substantial statistical errors. Additionally, the equation doesn't account for the fermionic nature of electrons, which requires antisymmetry when particles are exchanged. This requirement introduces nodes in the fermionic wavefunction; if not constrained, would lead to a bosonic solution. This issue is addressed by the fixed-node (FN) approximation\cite{Anderson}. This approximation constrains the wavefunction to maintain the nodal structure of a trial wavefunction, thereby introducing the fixed-node error as the sole source of error in DMC when the reference wavefunction is not exact. The accuracy of DMC thus heavily relies on the quality of the nodes in the trial wavefunction. By introducing a guiding or trial function, $\Psi_G\left(\textbf{R}\right)$, that closely approximates the ground state, the following transformation is applied:
\begin{equation}
f\left(\textbf{R},\tau\right) = \Psi_G\left(\textbf{R}\right)\Psi\left(\textbf{R},\tau\right),
\end{equation}
which modifies equation~(\ref{eq:Boson}) to:
\begin{align}\label{eq:Fermion}
-\frac{\delta f(\textbf{R},\tau)}{\delta \tau} &= \left[\sum_{i=1}^{N}-\frac{1}{2}\nabla^{2}_{i}f(\textbf{R},\tau) \right] - \nabla \cdot \left[\frac{\nabla \Psi(\textbf{R})}{\Psi(\textbf{R})}f(\textbf{R},\tau)\right] \\ \nonumber
&+ \left(E_L(\textbf{R})-E_T\right)f(\textbf{R},\tau),
\end{align}
$E_T$ is referred to as a "trial energy" and is used to keep the solution normalized over long-time scales, with $E_L(\textbf{R})$ representing the local energy at position $\textbf{R}$. The final term in Eq.~(\ref{eq:Fermion}) is a critical branching term that eliminates any 'walker' crossing a node (where the wavefunction changes sign) and duplicates any walker that reduces the system's energy, bringing it closer to the ground state. This mechanism is often described as the birth and death process in stochastic simulations.
\\
The accuracy of DMC hinges largely on the quality of the nodal surface defined by the underneath trial wavefunction. 
However, it is important to keep in mind that DMC is variational, meaning that the solutions we obtain are always an upper bound to the exact solution.\cite{ceperley84} This allows for the opportunity of testing with various guiding functions to identify the one that minimizes energy. 
For instance,  Bing et al.\cite{bing2023} demonstrated that using DFT with 3 different exchange-correlation (XC) functionals as guiding functions yielded consistent results within statistical errors for more than 1000 molecules from the QM5 dataset. 
Additionally, hybrid functionals can be employed to fine-tune the percentage of exact exchange, optimizing the energy further\cite{Busemeyer2016}. More complex trial wavefunctions, such as multi-Slater determinants generated from selected Configuration Interaction (sCI)\cite{Morales2012,caffarel2016,Scemama2018,malone2020systematic} or an orbital optimization paired with a variational Monte Carlo in the presence of a Jastrow factor\cite{slootman2024accurate}, can also be utilized to improve accuracy. These approaches improve the nodal surface  and therefore lower the FN-error associated to DMC, but often at a larger computational cost.
\\
\subsubsection{Computational details}
All 10,793 constitutional isomers (most stable conformer for each) containing up to 4 heavy atoms in VQM24 were selected for DMC calculations.
The total energy calculations were performed using the \texttt{QMCPACK} code~\cite{kim2018qmcpack,kent2020qmcpack}. 
For efficient sampling and to reduce
statistical fluctuations, we utilized a Slater-Jastrow type trial wavefunction for all DMC energy evaluations~\cite{Schmidt1990}:
\begin{equation}
\Psi_T(\vec{R}) = \exp\left[\sum_i J_i(\vec{R})\right]\sum_k^M C_kD_k^{\uparrow}(\varphi)D_k^{\downarrow}(\varphi),
\end{equation}
where $D_k^{\downarrow}(\varphi)$ denotes a Slater determinant composed of single-particle orbitals $\varphi_i=\sum^{N_b}_l C_l ^i \Phi_l$, in this study, constructed using PBE0\cite{PBE01,PBE02} Kohn-Sham (KS) orbitals as implemented in the \texttt{PySCF} code\cite{Pyscf2020}. 
Similarly to the study by Bing et al.\cite{bing2023}, using different functionals does not lead to very different nodal surfaces and results in energy differences of less than 1kcal/mol in DMC, for the small subset of molecules tested. To enhance efficiency and minimize fluctuations in regions close to ionic cores, ccECP pseudopotentials were applied to substitute core electrons\cite{ccecp_1,ccecp_2}.
These pseudopotentials, optimized for precise many-body methods like DMC, address non-local effects using the determinant-localization approximation and the t-moves strategy (DLTM).\cite{zen_new_2019, Casula2010} 
DMC evolving in real space shows in general minimal basis set size dependency, as documented in prior studies~\cite{Dubecky2017JCTC,malone2020systematic}. 
However, when the the basis set is chosen to be too small, the quality of the trial wavefunction can be severly affected degrading the quality of the nodal surface. Given that the cost of evaluating larger basis function is marginal in QMC, we ran all our simulations with the cc-pVQZ basis set, tailored for ccECP.
\\
The Jastrow function includes terms for one-body (electron-ion), two-body (electron-electron), and three-body (electron-electron-ion) interactions. The one- and two-body interactions were defined using spline functions\cite{Esler2012}, and the three-body interactions were modeled using polynomials.\cite{Drummond2004} Specifically, the study utilized 16 parameters for each atom type in one-body terms with a cutoff of 8 Bohr, and 20 parameters per spin-channel for two-body terms with a cutoff of 10 Bohr. 
The three-body terms incorporated 26 parameters each, with a 5 Bohr cutoff. These parameters in the Jastrow factor were individually optimized for each molecular geometry using a linear optimization method developed by Umrigar et al.\cite{umrigar07}. 
For all DMC simulations, we utilized a timestep of 0.001 a.u., eliminating the necessity for timestep extrapolation. 
The simulations involved 1500 blocks of 40 imaginary time steps each, with only the 40th step considered for calculating the standard deviation. 
We used 16,000 walkers to reduce autocorrelation and prevent population bias, achieving average error bars of 0.4 mHa across approximately 2.3 billion samples.
\\
Given the large number of molecules, we used the NEXUS Workflow package\cite{Nexus} to generate input files, manage, and monitor jobs across different stages of the calculations. This allowed for a "black box" and fully automated computational campaign. 
The trial wavefunction generation was conducted on the Argonne LCRC system, Improv, using a single node composed of 2x AMD EPYC 7713 with 64 cores at 2GHz. Each molecule, on average, required 45 seconds of computing, amounting to a total of 134 node hours for the whole set. The subsequent DMC calculations required 20 nodes per molecule on the Argonne Polaris HPC, using the AMD EPYC 7543P CPU with 64 cores at 2.8GHz. Each molecule took approximately 15 minutes to achieve a sub kcal/mol error bar, totaling around $\sim$54,000 node hours.
\\
\section{Data Records}
The dataset is available at Zenodo~\cite{vqm_dataset}.
The DFT dataset is reported as separate \texttt{.npz} files for, conformational minima, constitutional minima and saddle point structures.
Each property is reported in separate arrays with the ordering of the molecules across every array being the same.
The keys for accessing each property from the DFT \texttt{.npz} files are tabulated in Table 2.
\begin{table*}[!htb]
    \centering
    \renewcommand{\arraystretch}{0.5}
    \begin{tabular}{|l|c|l|}
        \hline
        \textbf{Property} & \textbf{Unit} & \textbf{Key} \\
        \hline
        Stoichiometry & - & \texttt{compounds}\\
        Atomic Numbers & - & \texttt{atoms}\\
        Cartesian coordinates (XYZ) & \AA &  \texttt{coordinates}\\
        SMILES & - & \texttt{graphs} \\
        InCHI strings & - & \texttt{inchi} \\
        Total energies & Ha & \texttt{Etot} \\
        Internal energies & Ha & \texttt{U0} \\
        Atomization energies & Ha & \texttt{Eatomization} \\
        Electron-electron energies & Ha & \texttt{Eee} \\
        Exchange correlation energies & Ha & \texttt{Exc} \\
        Dispersion energy & Ha & \texttt{Edisp} \\
        HOMO-LUMO gap & Ha & \texttt{gap} \\
        Dipole moments & a.u. & \texttt{dipole} \\
        Quadrupole moments & a.u. & \texttt{quadrupole} \\
        Octupole moments & a.u. & \texttt{octupole} \\
        Hexadecapole moments & a.u. & \texttt{hexadecapole} \\
        Rotational constants & MHz & \texttt{rots} \\
        Vibrational eigen modes & \AA & \texttt{vibmodes} \\
        Vibrational frequencies & cm$^{-1}$ & \texttt{freqs} \\
        Free energy (H) & Ha & \texttt{G} \\
        Internal (Thermal) energy (H) & Ha & \texttt{U298} \\
        Enthalpy (H) & Ha & \texttt{H} \\
        Zero point vibrational energy (H) & Ha & \texttt{zpves} \\
        Entropy (H) & $\frac{\textrm{cal}}{\textrm{mol K}}$ & \texttt{S} \\
        Heat capacities (H) & $\frac{\textrm{cal}}{\textrm{mol K}}$ & \texttt{Cv}, \texttt{Cp} \\
        Electrostatic potentials at nuclei & a.u. & \texttt{Vesp} \\
        Mulliken charges & a.u. & \texttt{Qmulliken} \\
        \hline
    \end{tabular}
    \caption{DFT properties for all the 835'947 converged molecules with up to 5 heavy atoms, along with the corresponding keys to access them from the reported \texttt{.npz} files in the VQM24 dataset.
    (H) indicates thermodynamic properties calculated via the Harmonic approximation.
    Molecular orbital energies are available in the wavefunction (\texttt{.molden}) files reported in a single tarball.}
    \label{dft_table}
\end{table*}
DMC data is similarly reported in a separate \texttt{.npz} file with the corresponding keys recorded in Table 3.
\begin{table*}[htb]
    \centering
    \renewcommand{\arraystretch}{1.2}
    \begin{tabular}{|l|c|l|}
        \hline
        \textbf{Property} & \textbf{Unit} & \textbf{Key} \\
        \hline
        Stoichiometry & - & \texttt{compounds}\\
        Atomic Numbers & - & \texttt{atoms}\\
        Cartesian coordinates (XYZ) & \AA &  \texttt{coordinates}\\
        SMILES & - & \texttt{graphs} \\
        InCHI strings & - & \texttt{inchi} \\
        Total energy & Ha & \texttt{Etot} \\
        Error bar & Ha & \texttt{std} \\
        \hline
    \end{tabular}
    \caption{DMC properties for 10'793  molecules with up to 4 heavy atoms, along with the corresponding keys to access them from the reported \texttt{.npz} file in the VQM24 dataset.}
    \label{dmc_table}
\end{table*}
All of the data, including the wavefunction files as a single tarball, is publicly available at the Zenodo repository~\cite{vqm_dataset}.
The wavefunction \verb|.molden| files also contain molecular orbital energies.

\section{Validation and analysis}
\begin{figure*}[!h]
    \centering
    \includegraphics[width=\linewidth]{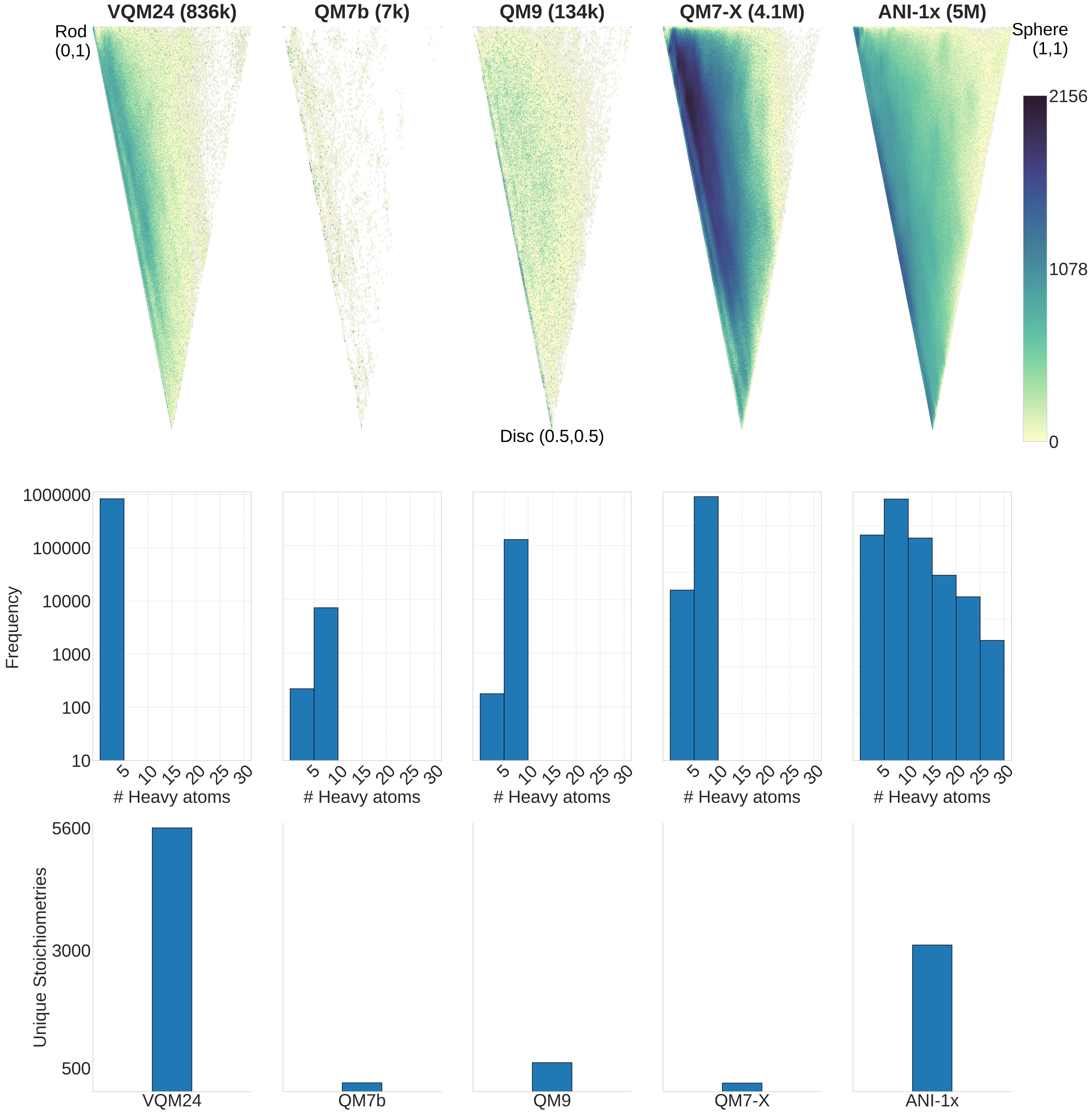}
    \caption{Comparison of structural diversities between VQM24 and 4 other commonly used QM datasets, QM7b~\cite{qm7b}, QM9~\cite{qm9}, QM7-X~\cite{qm7-x}, ANI-1x~\cite{ani1-x}.  
(A) (Top) Scatter plots of normalized ratios of principal moments of inertia (NPMI) for all molecules from the 5 datasets. Title includes number of molecules within each dataset in brackets with $k$ and $M$ indicating thousand and million respectively. Rod, disc and sphere indicate NPMI values corresponding to linear, flat and spherical systems respectively.  
(B) (Middle) Histograms binning molecules by the number of heavy atoms (non-Hydrogen).  
(C) (Bottom) Bar plots indicating number of unique stoichiometries.}
    \label{fig:vqm_fig4}
\end{figure*}
To analyze the structural diversity within the VQM24 dataset, we compared the distribution of  normalized ratios of principal moments of inertia (NPMI) values with 4 other QM (DFT) datasets containing molecules with similar sizes.
The corresponding scatter plots are shown in  fig.~\ref{fig:vqm_fig4}A.
Although VQM24 contains molecules of smaller size than the other 4 datasets (fig.~\ref{fig:vqm_fig4}B), it shows more comprehensive coverage of molecular shapes than the QM7b\cite{qm7b}, QM9\cite{qm9} datasets and a similar distribution to the larger QM7-X\cite{qm7-x} dataset (with lower density).
Due to the combinatorial sampling employed, the VQM24 dataset covers the 1-5 heavy atom chemical space more exhaustively than the other datasets as can be seen from fig.~\ref{fig:vqm_fig4}B and C.
Furthermore, due to the larger number of chemical elements included (10 for VQM24 compared to 5 for the others), greater chemical diversity can be expected within VQM24.
This leads to the much larger number of unique stoichiometries present in VQM24 when compared to QM7b/QM7-X and QM9 (fig.~\ref{fig:vqm_fig4}C).
\\
\begin{figure*}[!h]
    \centering
    \includegraphics[width=\linewidth]{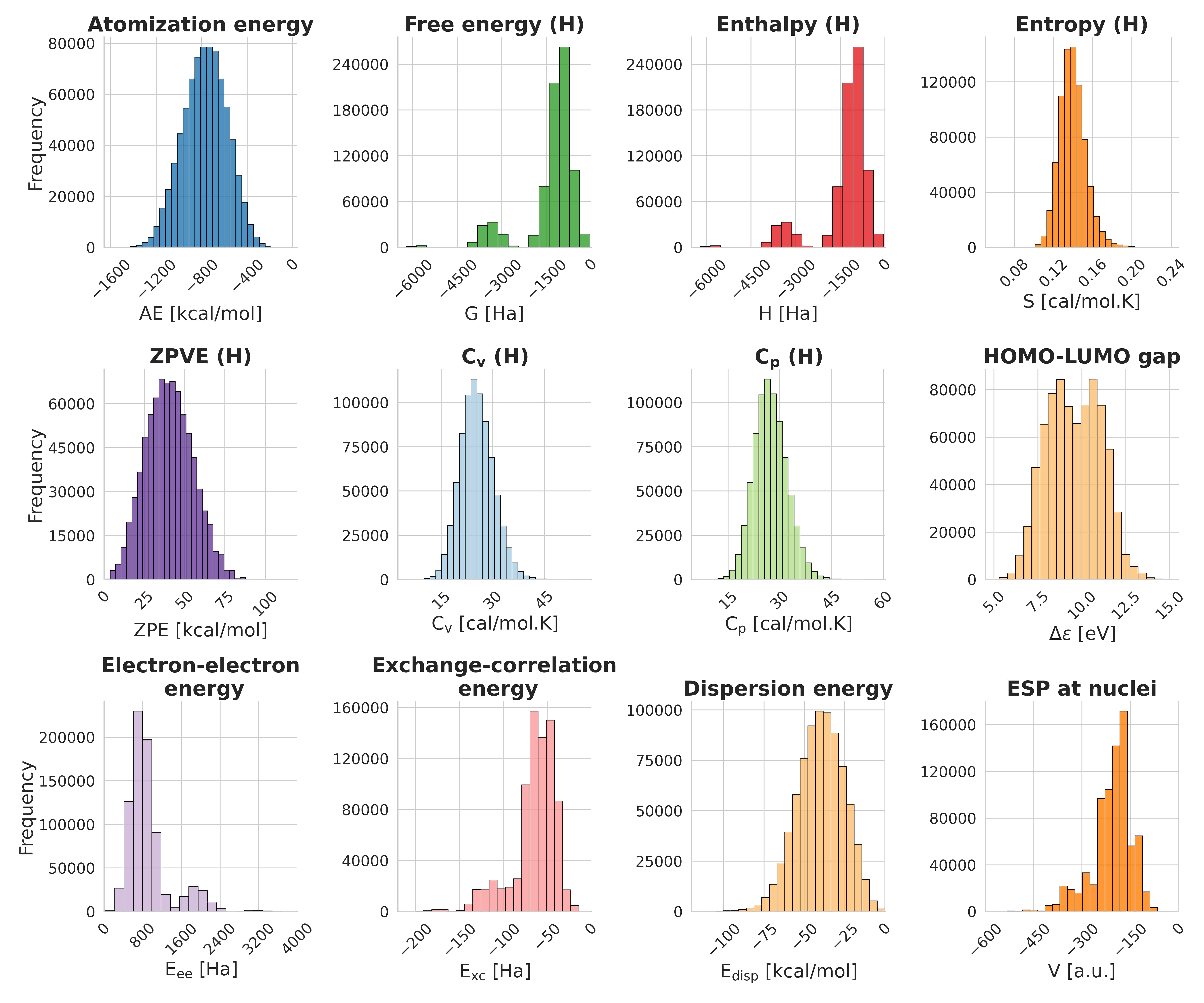}
    \caption{Distribution plots of 12 DFT ($\omega$B97X-D3/cc-pVDZ) calculated energetics out of the various properties reported in the VQM24 dataset.
(H) in the titles indicate thermodynamic quantities calculated via the harmonic approximation.
The electrostatic potential (ESP) at nuclei plot shows the sum of the ESP values at each nucleus within a molecule.
Units are mentioned in the x-axis labels.}
    \label{fig:vqm_fig5}
\end{figure*}
Energy ranges covered in VQM24 are shown in fig.~\ref{fig:vqm_fig5} which displays distibution plots of 12 properties derived from the total and electronic energies.
The atomization energies reported in VQM24 cover a range of 1545 kcal/mol.
\\
\begin{figure*}[!h]
    \centering
    \includegraphics[width=0.7\linewidth]{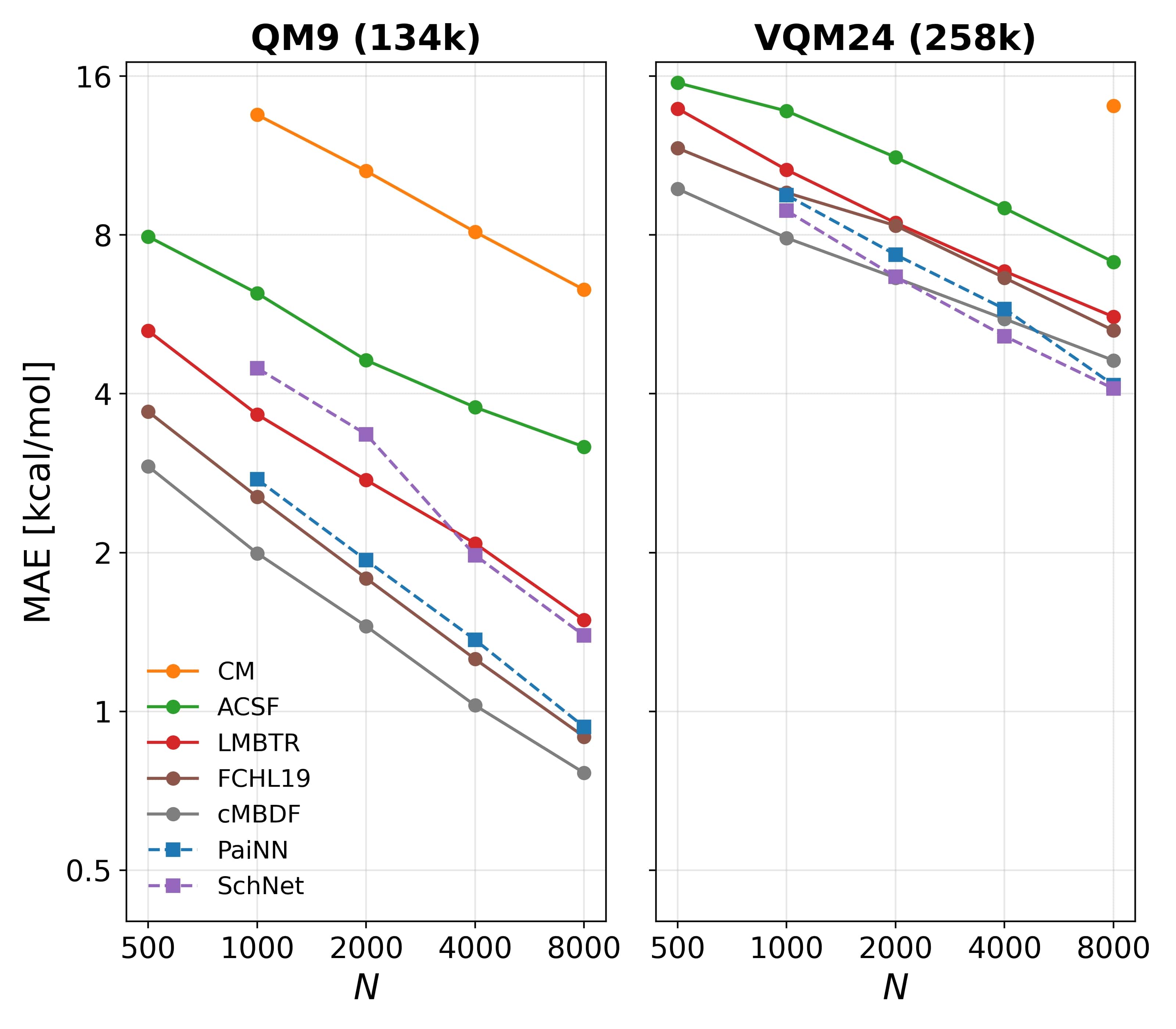}
    \caption{Atomization energy learning curves on the subset of 258k unique constitutional isomers
from VQM24 and the QM9 dataset.
Solid lines indicate ML models employing Kernel Ridge Regression (KRR) while dashed lines indicate Graph Neural Networks (GNN).
Representations used alongside KRR models are : Coulomb matrix~\cite{CM} (CM), atom-centered symmetry functions~\cite{acsf} (ACSF), many-body tensor representation~\cite{mbtr} (LMBTR), Faber-Christensen-Huang-Lilienfeld 19~\cite{fchl18,fchl19} (FCHL19), convolutional many-body distribution functionals~\cite{cmbdf,mbdf} (cMBDF/MBDF). 
GNNs employ the equivariant PaiNN~\cite{painn} and invariant SchNet~\cite{schnet} architectures.
Test set size in both cases was 10,000 randomly-selected molecules.
Plots show average of 5 such runs.}
    \label{fig:vqm_fig6}
\end{figure*}
To further assess the chemical diversity we trained and tested ML models for the task of predicting atomization energies within VQM24 and compare it to ML results obtained for the commonly used QM9 benchmark.
Fig. 6 presents the learning curves—i.e. the prediction error on atomization energies as a function of training‐set size—for both datasets.
We compare several kernel ridge regression (KRR)~\cite{vapnik1994learningcurves,gpr_rasmussen} models using different atomic representations, alongside invariant (SchNet~\cite{schnet}) and equivariant (PaiNN~\cite{painn}) message‐passing graph neural networks (GNNs).
The KRR and GNN models were trained and deployed using the {\tt QMLcode~\cite{qmlcode2017}} and {\tt Schnetpack~\cite{schnetpack}} libraries respectively.
Atomic gaussian kernel was used alongside all KRR models and the hyper-parameters (length-scale $l$, regularizer $\lambda$) were optimized via grid-search.
Logarithmic grids of $\left[0.1(2^{n}) ~ \forall~ n ~\in \{0,14\}\right]$ and $\left[10^{-3n}~ \forall~ n ~\in \{1,4\} \right]$ were employed for $l$ and $\lambda$ respectively.
Optimizations were performed via 5-fold cross-validation.
For both GNN models we employed the same hyper-parameters as in the PaiNN~\cite{painn} paper.
We note here that we used 128 atomic basis functions with both PaiNN and SchNet leading to a total parameter count of 589k for both models.
All models were trained for 1000 epochs using the Adam optimizer~\cite{adam} with a learning rate of $10^{-4}$.
Scripts for training and using these KRR and GNN models are available in the GitHub repository specified in the Code Availability section below.
\\
Since QM9 does not contain conformational isomers, Fig. 6B shows learning curves of atomization energies considering only the subset of the 258k lowest energy conformers (unique constitutional isomers) from VQM24 to be directly comparable.
While the range of atomization energies covered by VQM24 (1545 kcal/mol) is smaller than QM9 (2427 kcal/mol), evidently they are more challenging to learn as all ML models show upto $\sim$8 times larger mean errors than on QM9 for the same training set size.
This is likely due to the much larger chemical diversity of VQM24 which should make it a more challenging benchmark for the training and testing of ML models of the various chemically relevant physical properties reported.
\\
Using the best KRR model from Fig. 6 (cMBDF~\cite{mbdf,cmbdf}), we performed a prediction error analysis to detect outliers within the dataset.
We trained KRR/cMBDF based ML models on 200k molecules (4 disjointed training sets) to make predictions on the remaining dataset.
This was done on all the 784,875 equilibrium geometries (including conformers) reported in the dataset.
The resulting distribution of errors on the entire dataset is shown in fig. 7.
For each molecule we used the prediction with the smallest error out of the 3 ML models that were not trained on it.
The mean absolute error for the 784,875 predictions was obtained to be 0.75 kcal/mol with a standard deviation of 1.55 kcal/mol.
The largest obtained error was 167.3 kcal/mol and the 25 largest outliers show a mean absolute error of 85.9 kcal/mol.
These molecules are shown as insets in fig. 7.
\\
\begin{figure*}[!h]
    \centering
    \includegraphics[width=\linewidth]{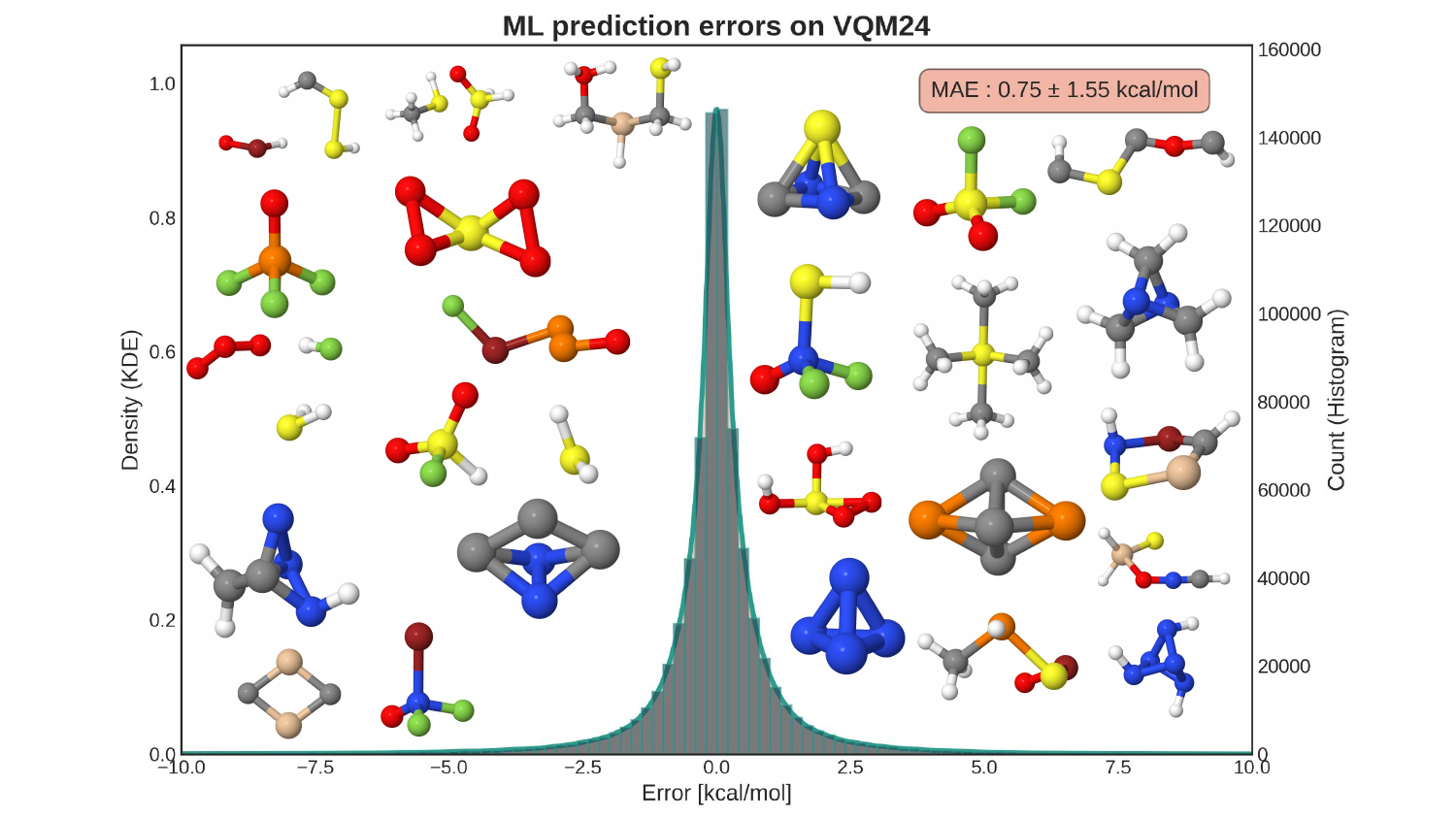}
    \caption{Atomization energy prediction errors on the 784,875 converged molecules from the VQM24 dataset.
All predictions were made by KRR models employing the cMBDF~\cite{cmbdf, mbdf} representation after training on 200k molecules.
Errors for all molecules were obtained by training four such models with disjointed training sets and using the smallest prediction error for all out-of-sample molecules.
Mean absolute error (MAE) for the predictions is shown in the top right inset along with the standard deviation.
Insets show molecules with the largest prediction errors.
Atomic colours correspond to: grey-Carbon, white-Hydrogen, blue-Nitrogen, red-Oxygen, dark green-Fluorine, cream-Silicon, orange-Phosphorous, yellow-Sulfur, green-Chlorine, dark red-Bromine.}
    \label{fig:vqm_fig7}
\end{figure*}

\section*{Code Availability}
Sample code for accessing the dataset is mentioned above.
Code used for generating the data and more tools can be found at 
\\
\verb|https://github.com/dkhan42/VQM24|

\section*{Acknowledgements}
We acknowledge the support of the Natural Sciences and Engineering Research Council of Canada (NSERC), [funding reference number RGPIN-2023-04853]. Cette recherche a été financée par le Conseil de recherches en sciences naturelles et en génie du Canada (CRSNG), [numéro de référence RGPIN-2023-04853].
This research was undertaken thanks in part to funding provided to the University of Toronto's Acceleration Consortium from the Canada First Research Excellence Fund,
grant number: CFREF-2022-00042.
O.A.v.L. has received support as the Ed Clark Chair of Advanced Materials and as a Canada CIFAR AI Chair.
O.A.v.L. has received funding from the European Research Council (ERC) under the European Union’s Horizon 2020 research and innovation programme (grant agreement No. 772834).
The authors are grateful to Compute Canada and the Acceleration Consortium for computational resources. 
A.B was funded by the U.S. Department of Energy, Office of Science, Basic Energy Sciences, Materials Sciences and Engineering Division, as part of the Computational
Materials Sciences Program and Center for Predictive Simulation of Functional Materials. 
DMC calculations used an award of computer time provided by the Innovative and Novel Computational Impact on Theory and Experiment (INCITE) program. 
This research used resources of the Argonne Leadership Computing Facility, which is a DOE Office of Science User Facility supported under contract DE-AC02-06CH11357. 
We also gratefully acknowledge the computing resources provided on IMPROV, a high-performance computing cluster operated by the Laboratory Computing Resource Center (LCRC) at Argonne National Laboratory.

\bibliographystyle{apsrev4-1}
\bibliography{sample.bib}
\end{document}